\documentstyle[manuscript,aps]{revtex}
\begin{document}
\draft
\title
{CHAOS {\sl vs} THERMALIZATION IN THE NUCLEAR SHELL MODEL}
\author{\bf Mihai Horoi$^{1,3}$, Vladimir Zelevinsky$^{1,2,4}$ {\rm and}
B. Alex Brown$^{1,2}$}
\address{$^{1}$National Superconducting Cyclotron Laboratory,
Michigan State University, \\
East Lansing, Michigan 48824-1321, USA\\
$^{2}$ Department of Physics and Astronomy, Michigan State University,\\
East Lansing, Michigan 48824-1116, USA\\
$^{3}$Institute of Atomic Physics, Bucharest, Romania\\
$^{4}$Budker Institute of Nuclear Physics, Novosibirsk 630090 Russia}
\maketitle

\begin{abstract}
Generic signatures of quantum chaos found in realistic shell model
calculations are compared with thermal statistical equilibrium.
We show the similarity of the informational entropy of individual
eigenfunctions in the mean field basis to the thermodynamical
entropy found from the level density. Mean occupation numbers
of single-particle orbitals agree with the Fermi-Dirac
distribution despite the strong nucleon interaction.

\end{abstract}


Chaotic dynamics is one of the most extensively developing
subjects in physics. Although classical deterministic chaos
is well understood \cite{LL,Schuster}
a rigorous definition of quantum chaos does not exist.
The relation to the classical limit is straightforward in problems of the
one-body type. Quantum billiards is the best known example. Classically,
regular or chaotic features of motion are determined by the shape of
boundaries. The corresponding
quantum level statistics \cite{Porter,Brody} display
local correlations and fluctuations of Poisson or Wigner type.
In many-body quantum systems a semiclassical
picture arises in the mean field approximation when
the system is modeled by a gas of independent quasiparticles
and symmetry (lack of symmetry) of the mean field determines
regularity (chaoticity) of single-particle motion.
As the excitation energy and level density increase,
the residual interactions between quasiparticles transform
the stationary states into exceedingly
complicated superpositions of the original "simple" configurations.
Already at early stages of this process the
local level statistics exhibit \cite{dro} features of chaos.
The pattern of chaotic signatures mixed with the apparent failure of the
independent quasiparticle model can be called "many-body chaos" \cite{Wies}.

In this paper we address the question of the relation between the
complicated structure of eigenstates and the general principles
of statistical mechanics. Having at our disposal exact eigenfunctions
of a model Fermi system with strong interactions (the nuclear shell model
\cite{Wild,shell}) we compare their statistical properties with those of the
equilibrium thermal ensemble.

The statistical approach implies that the observables
are insensitive to the actual microscopic state of
the system. Averaging over the equilibrium
ensemble should give the same outcome as an expectation
value for a typical single stationary wave function
at the same energy \cite{stat}. This requires the similarity of the
generic wave functions in a given energy region.
Perfect gases give the simplest example of many-body systems
where such properties of stationary states are evident.
To go beyond the perfect gas, we note that
the above description also fits the
notion of stochastic dynamics. In the classical case the
correspondence between statistical equilibrium and chaotic trajectories
exploring the whole energy surface is taken almost for
granted by many authors, see for example \cite{Ford}. As for the
quantum case, already the pioneering paper
on compound nucleus by Niels Bohr \cite{Niels} contains on
equal footing elements of both patterns, chaos and thermalization.
In discussing properties of chaotic wave functions, Percival
\cite{Perc} assumes that all of them "look
the same" and cover the entire available configuration space.
According to Berry \cite{Berry}, in systems with
the chaotic classical limit as a gas of hard spheres, the eigenfunctions
should behave like random superpositions of plane waves. This conjecture
is in fact equivalent to the microcanonical ensemble and
leads \cite{Mark} to the standard (Maxwell-Boltzmann,
Bose-Einstein or Fermi-Dirac) momentum distribution for
individual particles.

One can argue that the gas of hard spheres is a specific case of
many-body dynamics where the interaction is reduced to exclusion
of the inner volume of the spheres. However,
it was shown long ago by van Hove \cite{Hov} that the broader class of
gas-like systems displays quantum ergodicity: a random
initial wave function evolves with time into a
state which gives the same values of observables as the microcanonical
ensemble. The assumption of randomness or phase incoherence
is similar to Berry's conjecture or even to Boltzmann's
molecular chaos ({\sl Stosszahlansatz}).
Self-sustained Fermi systems like nuclei behave, according to
Landau-Migdal theory, analogously to the gas of
interacting quasiparticles. The residual interaction
cannot be reduced to rare pairwise collisions and is to be
treated on exact quantum-mechanical basis. The
generalization of the results derived for rigid spheres
to the strongly interacting case is not known. We address this question
by comparing the signatures of quantum chaos in the nuclear shell model
with the pattern of thermal equilibrium.

The actual computations were performed for 12 particles in the $sd$-shell
with the Wildenthal interaction \cite{Wild,oxba} which has been tested
earlier by numerous calculations of observables. Many-body basis states
$|k\rangle$ were constructed with good total angular momentum $J$,
its projection $M$, parity $\pi$ and isospin $T,T_{3}$.
Within this basis, the hamiltonian matrix has diagonal elements which
are dominated by the one-body part and numerically are spread over
the region from -120 to -60 MeV, and off-diagonal elements with an
average value of about 0.5 MeV. Eigenvalues
$E_{\alpha}$ for states with $J^{\pi}T$ equal to
$0^{+}0$ and $2^{+}0$ (with model space dimensions $N=839$ and $N=3273$
respectively) clearly showed chaotic level statistics \cite{shell}.
The amplitudes $C^{\alpha}_{k}$ of eigenfunctions
\begin{equation}
|J^{\pi}T;\alpha\rangle = \sum_{k}C^{\alpha}_{k}|J^{\pi}T;k\rangle  \label{1}
\end{equation}
have, for a given
$|\alpha\rangle$, Gaussian distribution with zero mean value and variance
$\overline{(C^{\alpha}_{k})^{2}} = 1/N^{\alpha}$. The
localization length $N^{\alpha}$ gives a measure of local chaos in the
vicinity of energy $E_{\alpha}$. In the extreme chaotic case $N^{\alpha}$
approaches the space dimension $N$ manifesting total mixing and
delocalization of eigenfunctions. We found $N^{\alpha}\approx 0.9N$ in
the middle of the spectrum for realistic interaction and
$N^{\alpha}\approx N$ for the degenerate model with no stabilizing influence
of the mean field. Instead of $N^{\alpha}$, it is also convenient to use
the informational entropy \cite{entr}
\begin{equation}
S^{\alpha} = - \sum_{k}(C^{\alpha}_{k})^{2}\ln[(C^{\alpha}_{k})^{2}] \label{2}
\end{equation}
as well as moments of the distribution function of the
components $C^{\alpha}_{k}$.
All such characteristics show that, as the excitation energy increases,
the eigenfunctions become more complex and the maximum of
complexity is reached in the middle of the spectrum. Our measures of
complexity are basis dependent. We argued \cite{shell} that the
mean field basis is preferential for such an analysis.

The same process of stochastization can be described in the
thermodynamic language. A closed equilibrated system with a sufficiently
high number of degrees of freedom is excited into an energy interval
($E,E+\Delta E$) where the density of states with given values of exact
integrals of motion ($J^{\pi}T$ in our case)
is $\rho(E)$. The average ("thermodynamical")
characteristics are determined by the statistical weight $\Omega(E)
= \rho(E)\Delta E$ the exact value of the uncertainty $\Delta E$ being not
important as far as $\Delta E \ll E$.
One can then define the thermodynamic entropy
$S^{th}(E) = \ln \Omega(E)$ and temperature $T$ according to
\begin{equation}
\frac{\partial S^{th}}{\partial E} = \frac{1}{T}.            \label{3}
\end{equation}
Of course, this description is basis-independent.

A system with a finite Hilbert space can be
heated until the level density saturates at maximum
entropy and infinite temperature (3). The local level density $\rho(E)$
for $N=839$ states $0^{+}0$ is presented as a histogram
in Fig. 1 together with a Gaussian fit
with the centroid at $E_{0}= - 90 MeV$ and variance $\sigma_{E} = 13$ MeV.
For such a fit,
the temperature (3) is $T= \sigma^{2}_{E}/(E_{0}-E)$. Similar results are
valid for other $J^{\pi}T$ classes and the Gaussian fit parameters turn out
to be the same. This means that one may speak about thermodynamic
equilibrium. The right half of the
spectrum is associated with decreasing entropy and negative temperature.

The Gaussian rather than semicircle $\rho(E)$ is expected
\cite{Brody} for a many-body system with two-body residual interaction.
The transition from semicircle to Gaussian level density occurs
\cite{Wong} when many-body forces are introduced
lifting the selection rules for interactions between configurations.
On the other hand, the banded random matrix theory predicts,
both numerically \cite{Casati} and analytically \cite{Kus},
the semicircle density for a sufficiently wide band of nonzero
matrix elements around the main diagonal. The realistic hamiltonian matrix
is banded in the basis of many-body configurations coupled via
two-body forces. But the matrix is far from being random since its
elements are linear combinations of only few (63 in the $sd$-shell)
two-body matrix elements.

To compare the global thermodynamic behavior with the properties
of individual eigenfunctions, we have calculated
the evolution of single-particle occupation numbers (the isoscalar monopole
component of the single-particle density matrix)
$n^{\alpha}_{\lambda}$ of the orbitals $\lambda = (l,j)$ along
the spectrum of stationary many-body states $|\alpha\rangle$, Eq.(1),
\begin{equation}
n^{\alpha}_{lj} = \frac{1}{2}\sum_{m\tau}\langle \alpha|a^{\dagger}_{ljm\tau}
a_{ljm\tau}|\alpha \rangle.                                 \label{4}
\end{equation}
The results are shown on Fig. 2 where the panels {\sl a,b} and {\sl c}
correspond to $0^{+}0, 2^{+}0$ and $9^{+}0 \quad (N=657$ states), respectively.
Although the states
within each $J^{\pi}T$ class are orthogonal and apparently have nothing
in common, all classes exhibit an identical smooth
behavior of occupation numbers. This is an additional
evidence for an equilibrated system.

In the center of the spectrum all
occupancies $f^{\alpha}_{lj} = n^{\alpha}_{lj}/(2j+1)$ become equal to
each other the common value being 1/2 for
our case of 12 particles in the $sd$-shell. It suggests that one
can associate to each eigenstate $|\alpha\rangle$ an effective
single-particle "temperature" $T^{\alpha}_{s-p}$ defined by the Fermi
distribution $f^{\alpha}_{lj}= \{\exp[(e_{lj}-\mu)/T^{\alpha}_{s-p}]+1\}^{-1}$.
$T^{\alpha}_{s-p}$ changes smoothly with $E_{\alpha}$ being
the same for all complicated wave functions near $E_{\alpha}$
as it should be for an intensive thermodynamic quantity. It becomes
infinite simultaneously with the thermodynamic temperature (3) when
the memory of the initial single-particle energies $e_{\lambda}$ is lost.

The microscopic mechanism of equilibration
can be understood from the analysis of fragmentation of
projected shell model states $|J^{\pi}T;k\rangle$. Applying
the recipes of statistical spectroscopy of French and Ratcliff \cite{FR}
one can explain the approximately constant
occupation of the $s_{1/2}$ orbital as well as the smooth evolution of
occupation factors for $d_{3/2}$ and $d_{5/2}$ orbitals as a function of
excitation energy. The conclusion is that the
thermodynamics of the system is determined mainly by the stabilizing
action of the mean field, despite the strong mixing of configurations.
An artificial reduction by a factor 10
of the diagonal matrix elements implies (Fig. 2{\sl d}) constancy of
occupation numbers, i.e. vanishing heat capacity. In this case one
has an analog of a dense hot gas with a very short quasiparticle
lifetime.

Using the occupancies $f^{\alpha}_{lj}$ of individual orbitals
one can calculate the single-particle entropy of
the quasiparticle gas \cite{stat} for each state $|\alpha \rangle$,
\begin{equation}
S^{\alpha}_{s-p} = - \sum_{lj}(2j+1)[f^{\alpha}_{lj}\ln f^{\alpha}_{lj}
+ (1 - f^{\alpha}_{lj})\ln (1 - f^{\alpha}_{lj})].               \label{5}
\end{equation}
Now we have three, apparently different, entropy-like
quantities: thermodynamic entropy $S^{th}(E) \sim \ln \rho(E)$, informational
entropy $S^{\alpha}$ (2) and single-particle entropy $S^{\alpha}_{s-p}$ (5),
the latter two for individual eigenstates. In Fig. 3 we juxtapose
the corresponding values of $\exp (S)$ for different physical situations,
I, II and III (columns). Rows {\sl a,b} and {\sl c} present the energy
behavior of $S^{th}, S^{\alpha}$ and $S^{\alpha}_{s-p}$, respectively,
for $0^{+}0$ states.

The column I of Fig. 3 shows the limit of a stable mean field and
relatively weak interaction (the diagonal matrix elements are amplified
by a factor 10). The thermodynamical entropy I{\sl a} is determined solely by
the level density. It displays Gaussian behavior of a pure combinatorial
nature typical for a slightly imperfect Fermi-gas in
a finite number of states.
It is quite similar to the single-particle picture
I{\sl c}. The informational entropy I{\sl b} is very low; only in the middle
region of enhanced level density does one see some effects of mixing.

The opposite case III corresponds to the strong residual interaction
(as in Fig. 2{\sl d} the diagonal matrix elements are reduced by a factor 10).
Almost all states are totally mixed and the informational
entropy III{\sl b} is near its chaotic maximum \cite{shell} of
$\exp(S^{\alpha})_{chaotic} = 0.48 N=404$ for this class of states.
$S^{\alpha}_{s-p}$ is also constant (III{\sl c}) on the maximum level
corresponding to the equiprobable population of orbitals.
Within the fluctuations, $S^{\alpha}$ and $S^{\alpha}_{s-p}$ coincide.
However, the level density still has the Gaussian shape (III{\sl a})
so that the system has normal thermodynamic properties.
It means that in the absence of the mean field the response of the
system to thermal excitation cannot
be formulated in terms of quasiparticle degrees of freedom. Therefore
the informational entropy calculated in the quasiparticle basis becomes
irrelevant from thermodynamic viewpoint.

The realistic case of the mean field consistent with the empirical
residual interaction is shown in column II.
When the magnitudes are normalized to each other, all three entropies
become identical within fluctuations. Therefore, using Eq.(3), we can identify
the temperature scales as well. For example, in Fig. 4 we show the
temperature which follows from Eq.(3) and Gaussian fit (Fig. 1)
for $\rho(E)$, as compared to $T^{\alpha}_{s-p}$ extracted from
the occupation factors of Fig. 2{\sl a-c}  with the use of the
Fermi distribution which
allows one to determine effective single-particle energies $e(lj)$. From the
horizontal line corresponding to the $s_{1/2}$ orbital, it follows that
the chemical potential $\mu \approx e(s_{1/2})$.
The $d$-orbitals can be determined
with respect to this level. In the $\alpha$-scale of Fig. 2, the occupation
numbers $n_{\lambda}$ have, except for the edges, a constant slope
$(dn_{\lambda}/d\alpha)_{0}$. Therefore
one can use the middle point of infinite temperature and the Gaussian width
$\sigma_{E}$ to find
\begin{equation}
e_{\lambda} - \mu = 4\rho(E_{0})\sigma_{E}^{2}\frac{1}{2j_{\lambda} + 1}
(dn_{\lambda}/d\alpha)_{0}.                                \label{6}
\end{equation}
Numerical values $e(d_{5/2})-\mu = - 3.4$ MeV and $e(d_{3/2})-\mu = 4.7$ MeV
for the effective single-particle energies
can be compared to the shell model
spin-orbit splitting 7.2 MeV near the ground state.

Thus, we have studied the relation of the individual properties of the
eigenstates to statistical features
of the equilibrium thermal ensemble. For the self-consistent
mean field and residual interactions, the thermodynamic entropy
defined either via the global level density or in terms of single-particle
occupation numbers behaves in the same way as the informational entropy
of generic compound states. It allows one to conclude that
(i) onset of "many-body" chaos is accompanied by the transition to
thermodynamic equilibrium; (ii) average equilibrium properties of a heated
many-body system with strong interactions can
still be described in terms of quasiparticles and their effective
energies in the appropriate mean field (this opens the way for explicit
calculation of matrix elements between compound states \cite{Fvorov}).

Let us stress the special role
of the mean field representation \cite{MF} both
for studying the degree of chaoticity of specific wave functions \cite{shell}
and for the statistical description. With the artificially
depressed or enhanced diagonal matrix elements, the level density
still keeps the Gaussian shape so the thermodynamic
entropy $S^{th}$ is qualitatively the same as in the realistic case
(Fig. 3{\sl a}). However, with no mean field (Fig. 3-III)
the increase of complexity
measured by the $S^{\alpha}$ and the mixing of quasiparticle configurations
measured by the $S^{\alpha}_{s-p}$, going together, are different from
the total heating measured by the level density and the "normal"
entropy $S^{th}$. The interaction is too strong and the
mixing does not depend on the actual level spacing. Almost
all wave functions "look the same" regardless of level density. It
means that, with no stable mean field, the quasiparticle "thermometer"
cannot resolve the spectral regions with different temperatures.

Finally we would like to give a more formal argument in favor of the
direct correspondence between chaos and thermalization. The general
description of a quantum system with non-complete information
uses the density matrix ${\cal D}$
which has, in an arbitrary many-body basis $|k\rangle$, matrix
elements ${\cal D}_{kk'} = \overline {C_{k}C^{\ast}_{k'}}$ where the
amplitudes are averaged \cite{stat} over the ensemble. If the ensemble
is generated by interaction with the environment, possible states
of the entire system are $|k;\nu\rangle$ where $\nu$ characterizes
the states of the environment compatible with the state $|k\rangle$ of the
subsystem under study. Then ${\cal D}_{kk'} = \sum_{\nu}C_{k\nu}
C^{\ast}_{k'\nu}$.
The corresponding statistical entropy $S = - Tr({\cal D}\ln{\cal D})$
is basis-independent and equals zero for pure states of the
isolated subsystem.
For canonical equilibrium ensembles $S$ coincides with the thermodynamic
entropy. Let us consider a gas of quasiparticles in
the ensemble generated by the residual interaction. This
makes sense only
after proper separation of global smooth dynamics from quasirandom
incoherent processes. Such a separation defines the optimal basis,
namely that of the self-consistent
mean field \cite{MF} (our "simple" states $|k\rangle$). Complicated stationary
states $|\alpha\rangle$ mimic the "total" system (quasiparticles + interaction
field). The ensemble average of ${\cal D}_{kk'} = \overline {C^{\alpha}_{k}
C^{\alpha}_{k'}}$
is to be taken over several neighboring states $|\alpha\rangle$.
If the amplitudes
$C^{\alpha}_{k}$ are uncorrelated and all neighboring states $|\alpha\rangle$
are similar, only diagonal elements of ${\cal D}_{kk'}$ survive and we
come to the informational entropy (2).

The authors would like to acknowledge support from the NSF grant 94-03666.
V.Z. is thankful to P.Cvitanovic and V.Sokolov for discussions.

\newpage
\begin{center}
{\bf Figure captions}
\end{center}
{\bf Figure 1}. Level density $\rho (E)$ for $0^{+}0$ states;
a histogram is compared with the Gaussian fit (dashed line).\\
\\
{\bf Figure 2}. Single-particle occupation numbers, Eq.(4), {\sl vs}
state number $\alpha$ for states
$0^{+}0$ (panel {\sl a}), $2^{+}0$ (panel {\sl b}), and $9^{+}0$ (panel
{\sl c}). Panel {\sl d} shows occupation numbers for $0^{+}0$ states
for diagonal matrix elements reduced by a factor 10.
For all panels the three curves (sets of points) refer
to $s_{1/2}, d_{3/2}$ and $d_{5/2}$ orbitals, from bottom to top. At
the middle of the spectrum all curves correspond to half-filled
orbitals (occupation numbers 1, 2 and 3 respectively).\\
\\
{\bf Figure 3}. \
Entropy-like quantities plotted as a function of energy for
$0^{+}0$ states. Columns correspond to the diagonal matrix elements
multiplied by factors of 10 (I), 1 (II) and 0.1 (III), the latter case
coincides with that of Fig. 2{\sl d}. Rows ${\sl a,b}$
and {\sl c} correspond to thermodynamic entropy,
informational entropy, Eq.(2), of individual states, and
single-particle entropy, Eq.(5), of individual states calculated from
the occupation numbers, respectively. The $e^{S}$
values in panels {\sl c} are
in units of $10^{4}$.\\
\\
{\bf Figure 4}. Temperature found from the Fermi-Dirac occupation numbers
of Fig. 2{\sl a}, points, and calculated from the global
fit to the level density of Fig. 1, solid line.

\end{document}